\documentclass[preprint,12pt,3p]{elsarticle}

\usepackage{graphicx}

\usepackage{amssymb}
\usepackage{amsthm}
\usepackage{bm}
\usepackage{amsmath}
\usepackage{hyperref}

\DeclareMathOperator*{\argmax}{arg\,max}

\journal{Physica A}

\begin{document}

\begin{frontmatter}

\title{Nonextensive triplets in stock market indices}

\author[addr1]{Dusan Stosic\corref{cor1}}
\address[addr1]{Centro de Inform\'atica, Universidade Federal de Pernambuco, Av. Luiz Freire s/n, 50670-901, Recife, PE, Brazil}
\cortext[cor1]{Corresponding author.}
\ead{dbstosic@bu.edu}

\author[addr1]{Darko Stosic}
\ead{ddstosic@bu.edu}

\author[addr2]{Tatijana Stosic}
\address[addr2]{Departamento de Estat\'{i}stica e Inform\'{a}tica, Universidade Federal Rural de Pernambuco, Rua Dom Manoel de Medeiros s/n, Dois Irm\~{a}os, 52171-900, Recife, PE, Brazil}
\ead{tastosic@gmail.com}

\begin{abstract}
Stock market indices are one of the most investigated complex systems in econophysics. Here we extend the existing literature on stock markets in connection with nonextensive statistical mechanics. We explore the nonextensivity of price volatilities for 34 major stock market indices between 2010 and 2019. We discover that stock markets follow nonextensive statistics regarding equilibrium, relaxation and sensitivity. We find nonextensive behavior in stock markets for developed countries, but not for developing countries. Distances between nonextensive triplets suggest that some stock markets might share similar nonextensive dynamics, while others are widely different. The current findings strongly indicate that the stock market represents a system whose physics is properly described by nonextensive statistical mechanics. Our results shed light on the complex nature of stock market indices, and establish another formal link with the nonextensive theory.


\end{abstract}


\end{frontmatter}


\section{Introduction}\label{secintro}


The stock market index is a good indicator of the overall market behavior and is frequently used by financial investors. Scale invariant behavior for both the distribution of returns and long term correlations in volatilities are well known properties of financial markets~\cite{Gopikrishnan,Liu}. While diverse methods have been employed to address the nonlinearity and complexity of these data, such as correlations~\cite{PlerouPRL,Plerou2000,Podobnik22079}, multifractal analysis~\cite{Zunino,Caraiani,Stosic2015}, network science~\cite{Onnela2004,Li,Marti}, and various entropy measures~\cite{Pincus13709,Zunino2010}, their complete understanding remains an open problem. The presence of fractal behavior~\cite{Caraiani}, long-range temporal dependencies~\cite{Liu} and heavy-tailed distributions~\cite{TsallisFinance}, indicate that stock markets deviate from the normal expectation into out-of-equilibrium states where traditional statistical mechanics does not work. A wide class of such systems can be described by nonextensive statistics~\cite{Tsallis} instead. The nonextensive theory thus forms a strong candidate for shedding new light on the phenomena involving stock market indices.

In the present study we consider a nonextensive theory which generalizes Boltzmann-Gibbs (BG) statistical mechanics for out-of-thermal equilibrium systems. From a mathematical standpoint, nonextensive statistics relies on a generalized definition of entropy~\cite{Tsallis}:
\begin{align}
S_q=-k\frac{1-\sum_ip_i^q}{1-q},
\end{align}
which is obtained by substituting exponentials with q-exponentials
\begin{align}
e_q(x)=[1+(1-q)x]^{1/(1-q)},
\end{align}
and natural logarithms with $q$-logarithms
\begin{align}
ln_q(x)=\frac{x^{1-q}-1}{1-q},
\end{align}
reducing to the BG entropy, the usual exponential and logarithm as $q\rightarrow 1$. The power law exponent $q$, also known as the entropic number, is intimately related to the microscopic dynamics and characterizes the degree of correlations in the system. A particularly important instance of nonextensive statistics is that of systems that find themselves out-of-thermal equilibrium but still form stationary states, which can be found in a great variety of complex systems. This seemingly simple generalization to physical systems implies that (i) stationary probability distributions acquire long tails, (ii) stationary states turn out less sensitive to initial conditions, and (iii) relaxation towards equilibrium becomes slower (q-exponential, rather than exponential). From a time series perspective, these three distinct properties translate into~\cite{Stosic}: (i) distributions with long (power law) tails, (ii) wider multifractal spectrum, and (iii) q-exponential decay of correlations. The set of obtained values for the nonextensive parameter $q$ is denoted as a ``q-triplet'' and characterizes metastable states in nonequilibrium. This triplet has been successfully explored in natural phenomena such as the ozone layer~\cite{Ferri}, solar plasma~\cite{Burlaga,Pavlos}, El Ni\~{n}o/Southern Oscillation~\cite{Rosso}, geological faults~\cite{deFreitas} and river discharge~\cite{Stosic}; in artificial systems including scale-free networks~\cite{TsallisPNAS}, logistic~\cite{Tirnakli} and standard~\cite{Ruiz} maps; and in financial systems namely cryptocurrencies~\cite{Stosic2018}.

The main aim of our work is to analyze the behavior of stock markets in the context of nonextensive formalism -- to compare their physical observables (such as price returns and volatilities) with those expected for a metastable dynamical system described by nonextensive statistics. In this context, we focus our attention on major stock market indices across the world from 2010 to 2019, in the aftermath of the last global financial crisis. Stock market volatilities are characterized by pronounced intermittency, exhibit unusually long temporal dependencies with correlations that decay slower than exponential, and tails in the probability distributions follow a power law with exponents less than three. We report evidence of nonextensive triplets~\cite{TsallisTriplet} that are characteristic of systems that follow nonextensive statistics, which implies that $q$-triplets are more common in finance than previously thought~\cite{Stosic2018}. The nonextensive behavior is mostly present in stock market indices from developed countries, but not from developing countries. Distances between the $q$-triplets further suggest that some stock markets might share similar nonextensive dynamics, while others are widely different. The discovery of $q$-triplets in stock markets, together with their previous appearance in cryptocurrencies~\cite{Stosic2018}, encourages further research into the prevalence of nonextensive phenomena in finance. 

The remainder of this paper is organized as follows: Section~\ref{secdata} describes the stock market data; Section~\ref{secres} presents the results and discussions; Section~\ref{secconcl} draws the conclusions.

\begin{figure}
\begin{center}
\includegraphics[width=\textwidth]{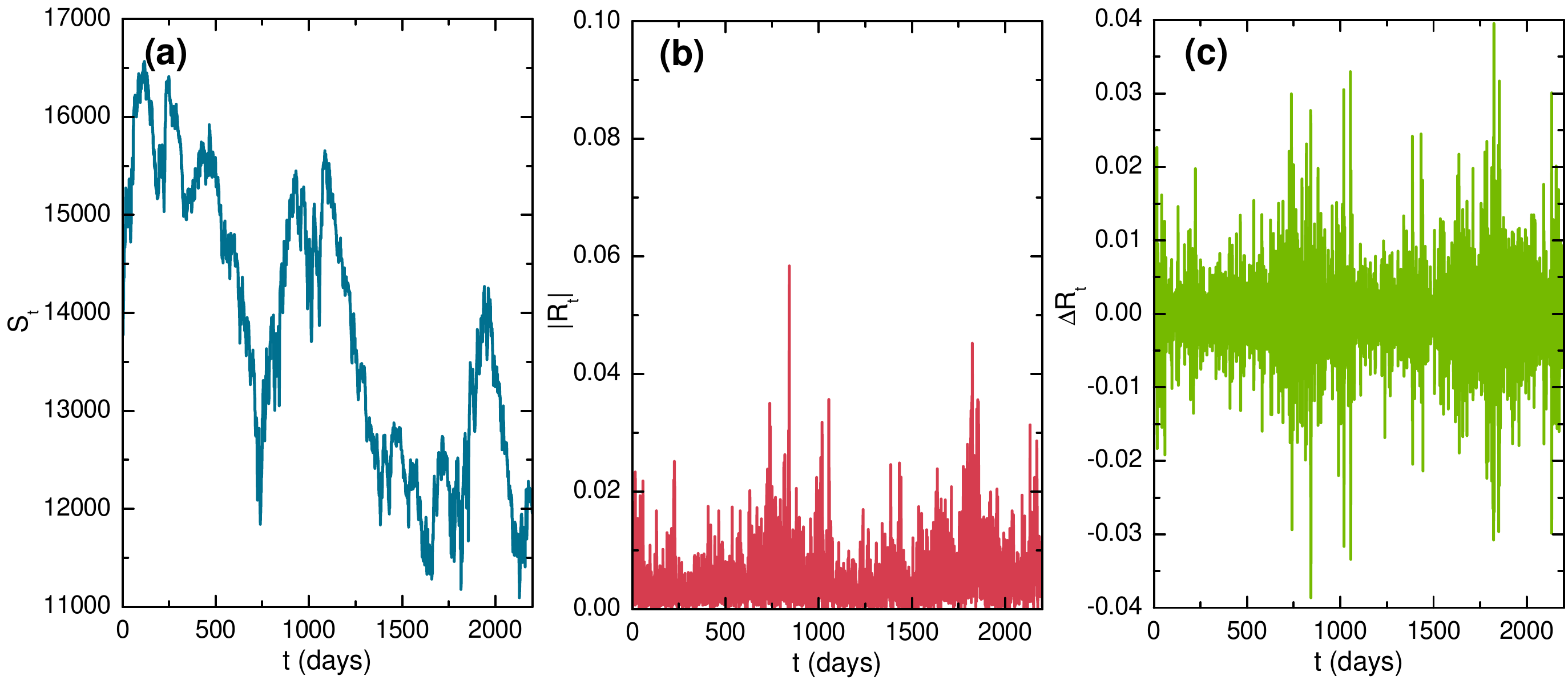}
\caption{
Time series of (a) closing prices $S_t$, (b) absolute daily logarithmic price changes $|R_t|$ (or volatility), and (c) volatility increments $\Delta R_t$ for the Canadian S\&P/TSX Composite (GSPTSE) market index.
\label{figdata}}
\end{center}
\end{figure}

\section{Data}\label{secdata}
We analyze the time series of $34$ major stock market indices that appear in the website~\url{https://www.investing.com/indices/major-indices} as listed in Table~\ref{tabdata}. The period under study starts from 2010, after the global financial crisis, and ends in 2019. For each of the stock markets we calculate the daily logarithmic change in closing price $S(t)$:
\begin{align}
R_t\equiv\ln\frac{S(t+1)}{S(t)},
\end{align}
and construct a time series from the volatilities $|R_t|$. We focus this investigation on the volatilities and their increments $\Delta R_t=|R_{t+1}|-|R_t|$. Fig.~\ref{figdata} reveals the presence of large price return and volatility variations with dense intermittent behavior which can be indicative of nonextensive statistics. 

\begin{table}[!t]
\caption{Information on analyzed time series for 34 major stock market indices.}
\label{tabdata}
\centering
\scalebox{0.75}{
\begin{tabular}{lllr}
\hline
Market & Country & Index & Period \\
\hline
AEX & Amsterdam & AEX & 1/4/2010 - 12/28/2018 \\
S\&P\/ASX 200 &	Australia & AXJO & 1/4/2010 - 12/28/2018 \\
BEL 20 & Germany & BFX & 1/4/2010 - 12/28/2018 \\
Budapest SE & Hungary & BUX & 3/7/2011 - 12/28/2018 \\
Bovespa	& Brazil & BVSP & 1/4/2010 - 12/28/2018 \\
Dow Jones Industrial Average & United States & DJI & 1/4/2010 - 12/28/2018 \\
CAC 40 & France & FCHI & 1/4/2010 - 12/28/2018 \\
FTSE MIB & Italy & FTMIB & 1/4/2010 - 12/28/2018 \\
FTSE 100 & England & FTSE & 1/4/2010 - 12/28/2018 \\
DAX & Germany & GDAXI & 1/4/2010 - 12/28/2018 \\
S\&P\/TSX Composite & Canada & GSPTSE & 1/4/2010 - 12/28/2018 \\
Hang Seng & Hong Kong & HIS & 11/4/2010 - 12/28/2018 \\
IBEX 35 & Spain & IBEX & 1/4/2010 - 12/28/2018 \\
MOEX Russian & Russia & IMOEX & 1/11/2010 - 12/28/2018 \\
NASDAQ Composite & United States & IXIC & 1/4/2010 - 12/28/2018 \\
Jakarta Stock Exchange Composite & Indonesia & JKSE & 1/4/2010 - 12/28/2018 \\
KOSPI	& South Korea & KS11 & 1/4/2010 - 12/28/2018 \\
Karachi 100 & Pakistan & KSE & 1/4/2010 - 12/28/2018 \\
S\&P\/BMV IPC & Mexico & MMX & 1/4/2010 - 12/28/2018 \\
Nikkei 225 & Japan & N225 & 1/4/2010 - 12/28/2018 \\
Nifty 50 & India & NSEI & 1/4/2010 - 12/28/2018 \\
OMX Stockholm 30 & Sweden & OMXS30 & 1/4/2010 - 12/28/2018 \\
PSEi Composite & Philippines & PSI & 11/2/2011 - 12/28/2018 \\
PSI 20 & Portugal & PSI20 & 5/25/2010 - 12/28/2018 \\
SET Index & Thailand & SETI & 3/18/2011 - 12/28/2018 \\
S\&P 500 & United States & SPX  & 1/4/2010 - 12/28/2018 \\
Shanghai Composite & China & SSEC  & 1/4/2010 - 12/28/2018 \\
SMI & Switzerland & SSMI  & 1/4/2010 - 12/28/2018 \\
FTSE Straits Times Singapore & Singapore & STI  & 3/7/2011 - 12/28/2018 \\
Euro Stoxx 50 & Euro Zone & STOXX50E  & 8/15/2011 - 12/28/2018 \\
TA 35 & Israel & TA35  & 1/3/2010 - 12/28/2018 \\
Tadawul All Share & Saudi Arabia & TASI & 1/2/2010 - 12/28/2018 \\
Taiwan Weighted & Taiwan & TWII & 3/17/2011 - 12/28/2018 \\
BIST 100 & Turkey & XU100 & 1/4/2010 - 12/28/2018 \\
\hline
\end{tabular}
}
\end{table}

\begin{figure}
\begin{center}
\includegraphics[width=\textwidth]{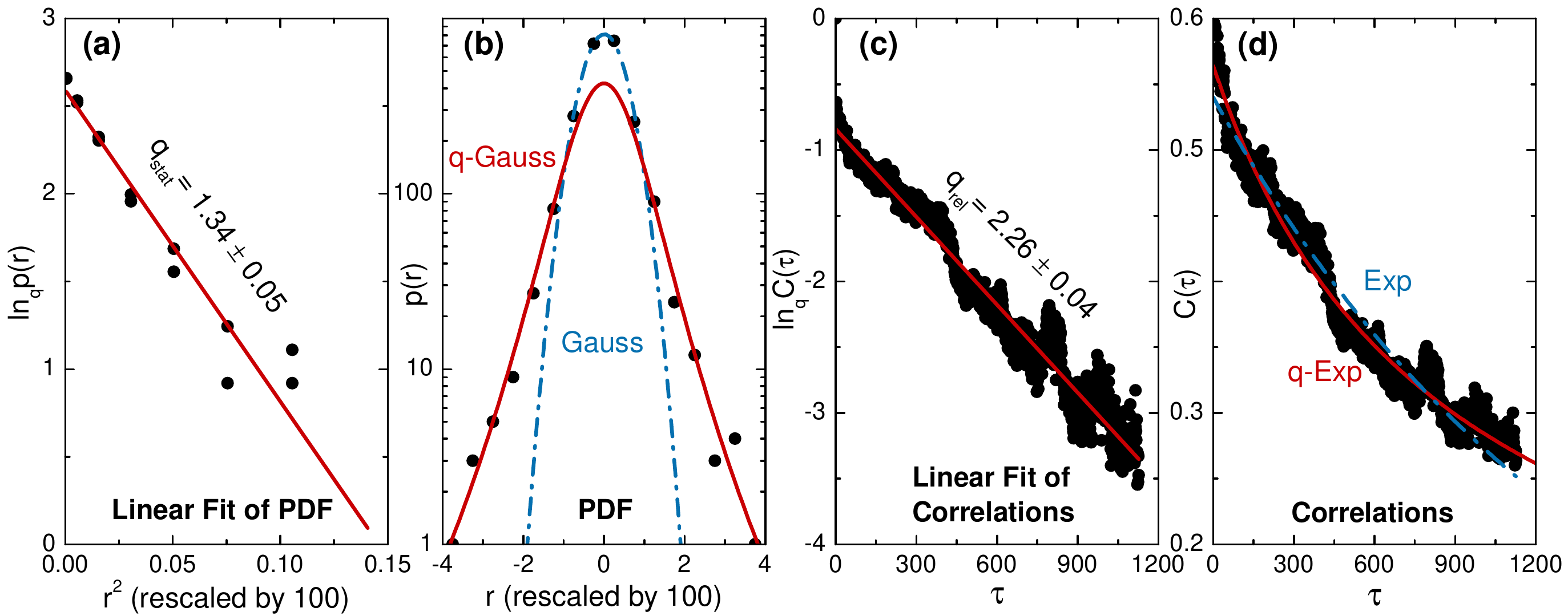}
\caption{
Fitting the $q_{\text{stat}}$ and $q_{\text{rel}}$ parameters to the Canadian market (GSPTSE). (a) Linear correlation between $\ln_q[p(r)]$ and $ r^2$. (b) Distribution of volatility increments $p(r)$ (circles), the $q$-Gaussian function that fits $p(r)$ (solid line), and the best fit with a standard Gaussian (dashed line). (c) Linear correlation between $\ln_q[C(\tau)]$ and  $\tau$. (d) Autocorrelation coefficient $C(\tau)$ \textit{vs.} time delay $\tau$ for volatility $|R|$ (circles), the $q$-exponential function that fits $C(\tau)$ (solid line), and the best fit with a standard exponential (dashed line).
\label{figgsptse}}
\end{center}
\end{figure}

\section{Results and discussion}\label{secres}
Systems that follows nonextensive statistics are characterized by a $q$-triplet $(q_{\text{stat}}, q_{\text{sens}}, q_{\text{rel}})\neq (1, 1, 1)$ that satisfies $q_{\text{stat}} > 1$, $q_{\text{sens}} < 1$ and $q_{\text{rel}} > 1$~\cite{TsallisTriplet}, or even the more rigid set of conditions $q_{\text{sens}}\leq 1\leq q_{\text{stat}}\leq q_{\text{rel}}$~\cite{TsallisBook}. Here we calculate nonextensive triplets for each stock market, where values of $q_{\text{stat}}$ are obtained from the $q$-Gaussian, $q_{\text{rel}}$ from the $q$-exponential, and $q_{\text{sens}}$ from the multifractality of the time series. We begin with the Canadian market (GSPTSE) as a case study and later expand to other stock markets. 

\subsection{Stationary $q=q_{\text{stat}}$}
A suitable $q$-value for the stationary state is obtained from the PDF associated to volatility increments $\Delta R_t=|R_{t+1}|-|R_t|$. $\Delta R$ range is subdivided into bins of width $\delta r$ centered at $r_i$ so we can obtain the frequency of $\Delta R$ values within each bin. The normalized histogram yields a stationary-PDF $\{p(r_i)\}^N_{i=1}$, where $p_i$ is the probability for a $\Delta R$ value to fall within the $i$th bin and $N$ is the number of bins. Our PDF is retrieved from adjusting the histogram of $\Delta R$ values to a $q$-Gaussian~\cite{Umarov}
\begin{align}
G_q(\beta;r) = \frac{\sqrt{\beta}}{C_q}e_q^{-\beta r^2},\qquad 1<q<3,\quad C_q=\frac{\sqrt{\pi}\Gamma\left(\frac{3-q}{2(q-1)}\right)}{\sqrt{q-1}\Gamma\left(\frac{1}{q-1}\right)},
\end{align}
in order to find the value of $q$ that best linearizes the graph $\ln_q[p(r_i)]$ vs. $r_i^2$. We vary $q$ from $0$ to $5$ and select which value makes the best linear adjustment from the coefficient of determination $R^2$~\cite{Pavlos}. We first consider volatility increments $\Delta R$ of the GSPTSE market index as a case study. The value $q_{\text{stat}}=1.34\pm 0.05$ is found to adjust well the experimental values with the coefficient of determination $R^2=0.968$ as shown in Fig.~\ref{figgsptse}(a). It should be emphasized that this $q_{stat}$ value is fully consistent with the bounds obtained from several independent studies involving the nonextensive framework (see, e.g.~\cite{TsallisFinance}). Fig.~\ref{figgsptse}(b) plots the associated $q$-Gaussian and the best adjustment that can be made with a (standard) Gaussian. Clearly the $p(r_i)$ values become noticeably non-Gaussian along the tails, and instead can be described by a power law. This is indicative of a Hamiltonian system whose elements do not interact locally but rather globally~\cite{TsallisTriplet}.


\begin{figure}
\begin{center}
\includegraphics[width=\textwidth]{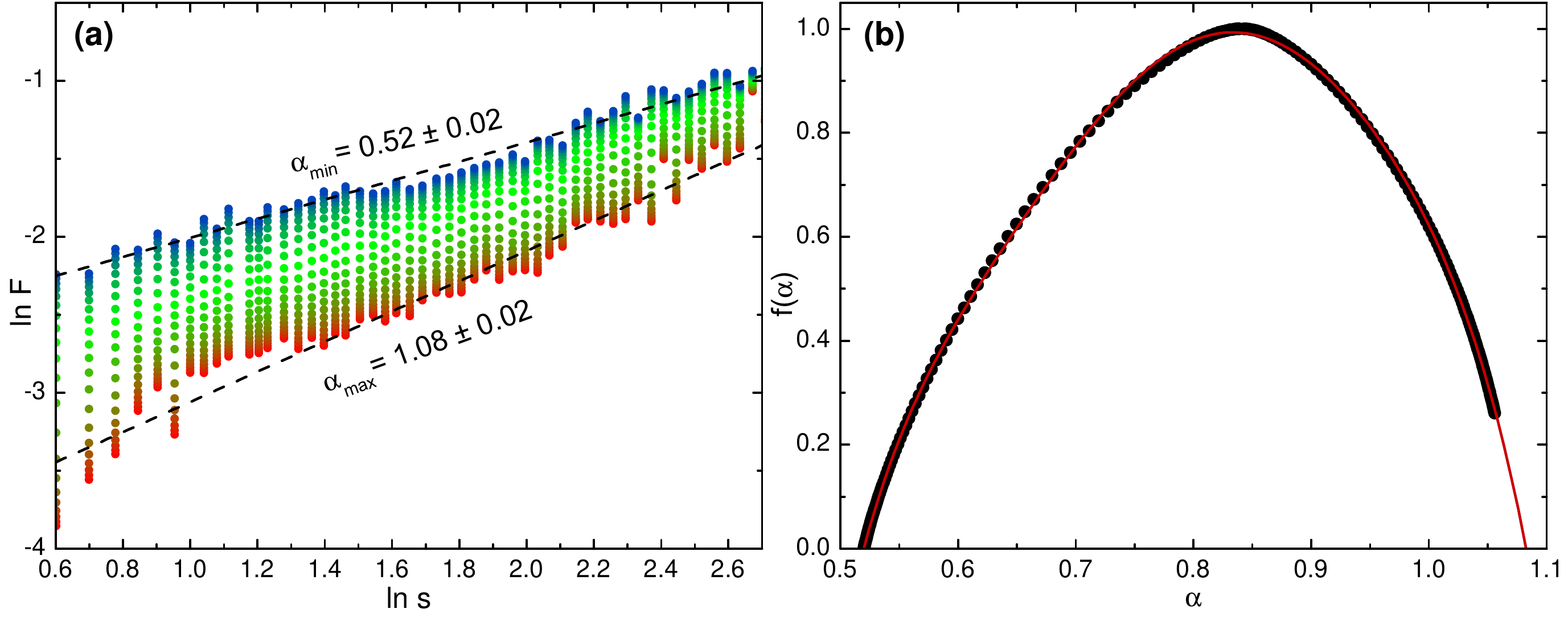}
\caption{
Fitting the $q_{\text{sens}}$ parameter to the Canadian market (GSPTSE). (a) Fluctuation function $\ln F(s)$ versus box size $\ln s$ (circles) and linear regressions for $\alpha_{\text{min}}$ and $\alpha_{\text{max}}$ (dashed lines). (b) Multifractal spectrum $f(\alpha)$ with a polynomial fit (solid line) where $q_{\text{sens}}=-2\times 10^{-4}\pm 0.02$.
\label{figqsens}}
\end{center}
\end{figure}

\subsection{Relaxation $q=q_{\text{rel}}$}
The corresponding $q_{rel}$ value, which describes a relaxation process, can be computed from the autocorrelation function
\begin{align}
C(\tau)=\frac{\sum_t |R_{t+\tau}|\cdot |R_t|}{\sum_t |R_t|^2}.
\end{align}
For a classical BG process such correlation should decay in exponential fashion, but volatilities in stock markets are known to exhibit long-range correlations~\cite{TsallisFinance}. Fig.~\ref{figgsptse}(d) clearly shows that the autocorrelation of the series $|R_t|$ decays much slower than an exponential function. We can estimate the value of $q_{rel}$ by the same linear adjustments on the graph $\ln_q[C(\tau)]$ vs. $\tau$ to determine which choice of $q$ best linearizes the data. Fig.~\ref{figgsptse}(c) reveals that autocorrelations of GSPTSE market index decay as a power law with an exponent $q_{\text{rel}}=2.26\pm 0.04$ and $R^2=0.965$. This suggests that macroscopic variables of the considered stock market decay slower than exponential to their equilibrium values.


\subsection{Sensitivity to initial conditions $q=q_{\text{sens}}$}
Systems well-described by the nonextensive theory exhibit less than exponential sensitivity to initial conditions. Small initial differences between neighboring states grow in $q$-exponential fashion characterized by a vanishing Lyapunov exponent and a parameter $q_{sens}$. Deviations of the neighboring trajectories of the attractor set of the dynamics leads to a multifractal structuring of the phase space~\cite{Pavlos}. The $q_{sens}$ value can be derived from the multifractal spectrum $f(\alpha)$ of the attractor associated to nonlinear dynamical system, reflected by $|R_t|$. $f(\alpha)$ denotes the fractal dimension of the attractor's subset that possesses the local scaling exponent $\alpha$~\cite{Kantelhardt}. The extreme values $\alpha_{min}$ and $\alpha_{max}$ of the multifractal spectrum, for which $f(\alpha) = 0$, have the following relation to the $q_{\text{sens}}$ parameter~\cite{Lyra}:
\begin{align}
\frac{1}{1-q_{\text{sens}}}=\frac{1}{\alpha_{\text{min}}}-\frac{1}{\alpha_{\text{max}}}.
\end{align}
We calculate $f(\alpha)$ using the Multifractal Detrended Fluctuation Analysis (MFDFA) method which is superior to other methods when dealing with non-stationary time series~\cite{Kantelhardt}. Fig.~\ref{figqsens}(a) shows that the fluctuation function $F_{\eta}(s)$ increases with the box size $s$ as a power law $F_{\eta}(s)\sim s^{h(\eta)}$, where the scaling exponent $h(\eta)$ is calculated as the slope of the linear regression of $\ln F_{\eta}(s)$ \textit{vs.} $\ln s$. These exponents are related to the singularity spectrum $f(\alpha)$ through a Legendre transformation~\cite{Kantelhardt} from which the spectra extrema can be obtained by extrapolating the curve of a polynomial fit to zero. Fig.~\ref{figqsens}(b) reveals that the GSPTSE spectrum has a wide range of scaling exponents $\Delta\alpha=\alpha_{max}-\alpha_{min}=0.56$ and the position of maximum $\alpha_0=\argmax f(\alpha)=0.83$ is consistent with values obtained in literature for volatilities in stock markets~\cite{JIANG2008}. From the spectrum extrapolation we obtain $\alpha_{\text{min}}=0.52\pm 0.02$ and $\alpha_{\text{max}}=1.08\pm 0.02$ resulting in the parameter $q_{\text{sens}}=-2\times 10^{-4}\pm 0.02$. This low value for $q_{\text{sens}}$ indicates that its distribution exhibits weak chaos~\cite{Tsallis} in the full dynamical space of the system~\cite{Tsallis,Burlaga}.

\subsection{$q$-triplets for stock markets}
Our analysis so far reveals a $q$-triplet for the Canadian market (GSPTSE) that obeys the general relation $q_{\text{sens}}\leq 1\leq q_{\text{stat}}\leq q_{\text{rel}}$~\cite{TsallisBook}, and is consistent with the nonextensive scenario. The question remains whether similar relations appear for other stock market $q$-triplets. 
Out of 34 analyzed stock market indices, we find that 22 stock markets satisfy the aforementioned $q$-triplet relation, as underlined in Tab.~\ref{tabtriplet}. Interestingly most of these stock markets come from developed countries, while developing countries such as Brazil (BVSP) and Russia (IMOEX) tend to escape from the nonextensive formalism. This suggests that stock markets from developed countries such as United States (SPX and DJI) may well represent a system in an off-equilibrium stationary state whose physics is properly described by nonextensive statistical mechanics. For developing countries the stock market indices often reverse the relationship between $q_{stat}$ and $q_{rel}$ such that $q_{stat}>q_{rel}$. Ref.~\cite{Rosso} observed a similar inversion in the $q$-triplet relation for variability of El Ni\~no oscillations.

\begin{figure}
\begin{center}
\includegraphics[width=\textwidth,trim={3cm 12.5cm 0 0},clip]{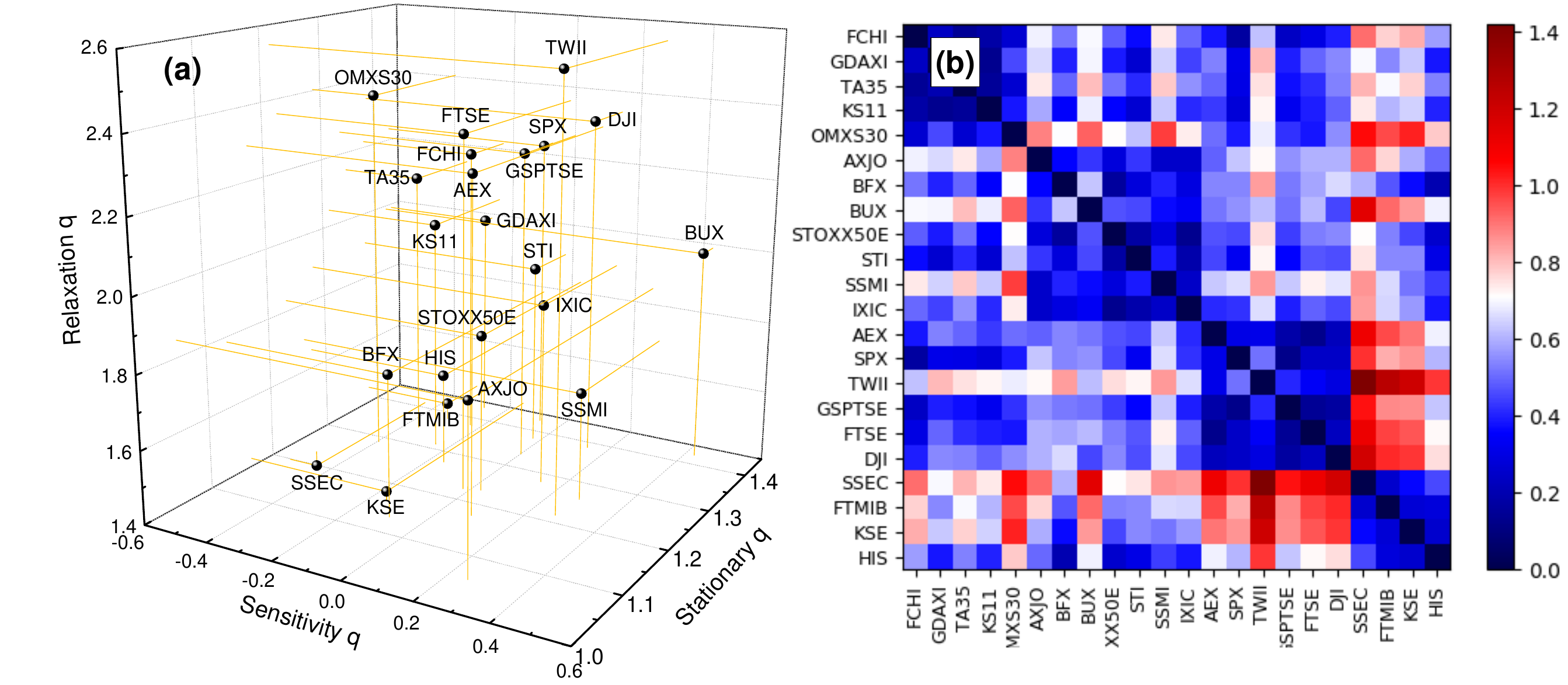}
\caption{
(a) Scatter plot of the nonextensive triplet values for the $22$ stock markets that are nonextensive. (b) The distance matrix $D$ containing their pairwise distances $d(i,j)$. 
\label{figtriplet}}
\end{center}
\end{figure}

The wide range of $q$-triplets observed earlier suggests the presence of distinct nonextensive dynamical behavior between stock markets -- a stationary or metastable state is characterized by a triplet of $q$-values. This motivates for a comparison between the set of $q$-values in order to obtain useful insight on the different dynamical processes between stock market indices. We map $q$-triplets to the three-dimensional Cartesian space in Fig.~\ref{figtriplet}(a), where we place different $q$-values on the same scale. We discover that certain stock markets form clusters such as on the bottom left (AXJO, BFX, HIS and FTMIB) and upper middle (AEX, FCHI, FTSE, GSPTSE, SPX, TA35) of the space. Other stock markets such as BUX, KSE, OMXS30, SSEC, TWII remain isolated and well separated from the clusters.

To better quantify the differences stated above, we define a distance metric between $q$-triplet pairs as follows~\cite{Stosic2018}:
\begin{align}
d(a,b)=\sqrt{(q^a_{\text{sens}}-q^b_{\text{sens}})^2+(q^a_{\text{stat}}-q^b_{\text{stat}})^2+(q^a_{\text{rel}}-q^b_{\text{rel}})^2},
\end{align}
where the $q$-values are indexed by $a$ or $b$ for the given pair. We can then construct a distance matrix $D_{ij}=d(i,j)$ containing pairwise distances between the 22 stock market $q$-triplets that follow nonextensive statistics. By construction the elements $D_{ij}=0$ correspond to stock market pairs that share the same $q$-triplet, and perhaps the same underlying dynamical process. The rows and columns (or stock market pairs) of the distance matrix $D$ are grouped using a spectral block clustering procedure~\cite{Dhillon}. Fig.~\ref{figtriplet}(b) reveals that a complex arrangement of $q$-triplet distances ($D_{ij}$) exists between the different stock markets. The matrix consists of blocks of small distances which represent clusters of nearby $q$-triplet values. We also discover that certain pairs of stock market $q$-triplets are separated by a large distance $d(i,j)$. For example, Taiwan (TWII) is very distant from other Asian stock markets (SSEC, KSE, HSI). China (SSEC) is distant from markets in the Americas (DJI, GSPTSE, SPX), but close to many markets in Asia (KSE and HSI). Sweden (OMXS30) is distant from Asian markets (HSI, KSE, SSEC, TWII) and also from the Euro market (STOXX50E). Many other similar conclusions can be drawn from Fig.~\ref{figtriplet}(b). These nontrivial distances could indicate similarities in nonextensive behavior of the underlying processes that characterize stock market indices.


\begin{table}[!t]
\caption{Nonextensive triplets for $34$ major stock market indices.}
\label{tabtriplet}
\centering
\scalebox{0.75}{
\begin{tabular}{|l|r|r|r||l|r|r|r|}
\hline
Market Index & $q_{\text{sens}}$ & $q_{\text{stat}}$ & $q_{\text{rel}}$ & Market Index & $q_{\text{sens}}$ & $q_{\text{stat}}$ & $q_{\text{rel}}$\\
\hline
\underline{AEX}	& $	0.08	\pm	0.04	$ & $	1.18	\pm	0.09	$ & $	2.30	\pm	0.04	$ &	\underline{KSE}	& $	-0.16	\pm	0.02	$ & $	1.16	\pm	0.06	$ & $	1.42	\pm	0.02	$	\\
\underline{AXJO}	& $	0.25	\pm	0.03	$ & $	1.06	\pm	0.11	$ & $	1.85	\pm	0.02	$ &	MMX	& $	0.25	\pm	0.02	$ & $	1.55	\pm	0.15	$ & $	1.01	\pm	0.01	$	\\
\underline{BFX}	& $	-0.09	\pm	0.02	$ & $	1.13	\pm	0.09	$ & $	1.79	\pm	0.02	$ &	N225	& $	0.62	\pm	0.04	$ & $	1.63	\pm	0.11	$ & $	0.64	\pm	0.01	$	\\
\underline{BUX}	& $	0.43	\pm	0.03	$ & $	1.42	\pm	0.06	$ & $	1.99	\pm	0.03	$ &	NSEI	& $	0.81	\pm	0.02	$ & $	1.19	\pm	0.11	$ & $	0.96	\pm	0.01	$	\\
BVSP	& $	0.12	\pm	0.02	$ & $	1.33	\pm	0.03	$ & $	0.80	\pm	0.00	$ &	\underline{OMXS30}	& $	-0.40	\pm	0.02	$ & $	1.29	\pm	0.10	$ & $	2.40	\pm	0.04	$	\\
\underline{DJI}	& $	0.14	\pm	0.03	$ & $	1.39	\pm	0.08	$ & $	2.34	\pm	0.03	$ &	PSI	& $	-0.01	\pm	0.03	$ & $	1.50	\pm	0.08	$ & $	1.46	\pm	0.02	$	\\
\underline{FCHI}	& $	-0.23	\pm	0.03	$ & $	1.38	\pm	0.05	$ & $	2.22	\pm	0.03	$ &	PSI20	& $	-0.03	\pm	0.03	$ & $	1.35	\pm	0.08	$ & $	0.66	\pm	0.01	$	\\
\underline{FTMIB}	& $	-0.32	\pm	0.03	$ & $	1.38	\pm	0.03	$ & $	1.45	\pm	0.01	$ &	SETI	& $	0.24	\pm	0.02	$ & $	1.67	\pm	0.13	$ & $	1.23	\pm	0.02	$	\\
\underline{FTSE}	& $	-0.02	\pm	0.02	$ & $	1.23	\pm	0.12	$ & $	2.36	\pm	0.04	$ &	\underline{SPX}	& $	-0.07	\pm	0.03	$ & $	1.43	\pm	0.07	$ & $	2.23	\pm	0.03	$	\\
\underline{GDAXI}	& $	-0.27	\pm	0.03	$ & $	1.44	\pm	0.05	$ & $	1.99	\pm	0.02	$ &	\underline{SSEC}	& $	-0.51	\pm	0.03	$ & $	1.23	\pm	0.06	$ & $	1.36	\pm	0.02	$	\\
\underline{GSPTSE}	& $	0.00	\pm	0.02	$ & $	1.34	\pm	0.06	$ & $	2.26	\pm	0.04	$ &	\underline{SSMI}	& $	0.28	\pm	0.03	$ & $	1.26	\pm	0.07	$ & $	1.70	\pm	0.02	$	\\
\underline{HIS}	& $	-0.16	\pm	0.02	$ & $	1.27	\pm	0.05	$ & $	1.65	\pm	0.02	$ &	\underline{STI}	& $	-0.02	\pm	0.03	$ & $	1.38	\pm	0.10	$ & $	1.91	\pm	0.03	$	\\
IBEX	& $	-0.06	\pm	0.03	$ & $	1.39	\pm	0.07	$ & $	1.20	\pm	0.01	$ &	\underline{STOXX50E}	& $	0.03	\pm	0.02	$ & $	1.23	\pm	0.05	$ & $	1.83	\pm	0.03	$	\\
IMOEX	& $	0.26	\pm	0.02	$ & $	1.28	\pm	0.04	$ & $	1.29	\pm	0.01	$ &	\underline{TA35}	& $	-0.36	\pm	0.03	$ & $	1.35	\pm	0.12	$ & $	2.15	\pm	0.04	$	\\
\underline{IXIC}	& $	0.14	\pm	0.02	$ & $	1.28	\pm	0.08	$ & $	1.90	\pm	0.02	$ &	TASI	& $	-0.04	\pm	0.02	$ & $	1.35	\pm	0.11	$ & $	1.19	\pm	0.01	$	\\
JKSE	& $	0.08	\pm	0.03	$ & $	1.21	\pm	0.07	$ & $	1.04	\pm	0.01	$ &	\underline{TWII}	& $	0.27	\pm	0.02	$ & $	1.22	\pm	0.10	$ & $	2.55	\pm	0.05	$	\\
\underline{KS11}	& $	-0.25	\pm	0.03	$ & $	1.31	\pm	0.11	$ & $	2.05	\pm	0.03	$ &	XU100	& $	0.20	\pm	0.02	$ & $	1.28	\pm	0.05	$ & $	1.05	\pm	0.01	$	\\
\hline
\end{tabular}
}
\end{table}

\section{Conclusions}\label{secconcl}
In this work we study the nonextensive behavior of daily volatilities for major stock market indices from $2010$ to $2019$. We find that for many developed countries the corresponding $q$-triplets satisfy the relation $q_{\text{sens}}\leq 1\leq q_{\text{stat}}\leq q_{\text{rel}}$~\cite{TsallisBook}. For developing countries the relationship between $q_{stat}$ and $q_{rel}$ reverses and so they cannot be explained by the nonextensive framework. Distances between $q$-triplets are such that many stock markets fall near each other after mapping them to a three-dimensional space. This could be indicative of similar nonextensive dynamics. Some markets are separated by a large distance which implies distinct nonextensive behavior. While not every stock market index may be expected to follow nonextensive statistics, our particular case demonstrates that intermittency, slowly decaying correlations and heavy tailed distributions may be well explained through this novel theory. It remains to be seen whether the proximity of $q$-triplets means similar mechanisms are responsible for stock market dynamics. The current findings suggest that further research should be directed at searching for signs of nonextensivity in other financial systems.

\bibliographystyle{elsarticle-num}

\bibliography{tripletmarket.bbl}

\end{document}